\documentclass[pra,showpacs]{revtex4}
\usepackage{amssymb}
\usepackage{graphicx}
\usepackage{dcolumn}
\usepackage{amsmath}

\begin{document}
\title{Quantum Hamilton-Jacobi Approach to Two Dimensional
Singular Oscillator}
\date{\today}
\author{\"{O}zlem Ye\d{s}ilta\d{s} $^1$}
\email{yesiltas@gazi.edu.tr}
\author{Beng\"{u }Demircio\u glu $^2$}
\email{bengudemircioglu@yahoo.com}

\affiliation{ $1$ Gazi University, Faculty of Arts and Sciences,
Physics Department, 06500 Teknikokullar, Ankara, Turkey\\
$2$ Ankara University, Faculty of Sciences,\\ Department of Physics,
06100 Tando\u gan, Ankara, Turkey}

\begin{abstract}
\noindent  We have obtained the solutions of two dimensional
singular oscillator which is known as the quantum
Calogero-Sutherland model both in cartesian and parabolic
coordinates within the framework of quantum Hamilton Jacobi
formalism. Solvability conditions and eigenfunctions are obtained by
using the singularity structures of quantum momentum functions under
some conditions. New potentials are generated by using the first two
states of singular oscillator for parabolic coordinates.
\end{abstract}


\maketitle

In quantum mechanics, there are only limited number of exactly
solvable (ES) systems such that their eigenvalues and eigenfunctions
can be obtained in an explicit and closed form [1,2,3,4,5]. Another
definition is proposed that a quantum mechanical system is called as
exactly solvable if the solutions of Schr\"{o}dinger equation can be
expressed in terms of hypergeometric functions $_m F_n $ \cite{6}.
There are some well-known potentials such as harmonic oscillator,
Morse, trigonometric/modified P\"{o}schl-Teller,
trigonometric/hyperbolic Manning-Rosen, and the Natanzon potentials
\cite{7,8,9} which are one-dimensional exactly solvable systems and
their eigenfunctions can be stated in terms of hypergeometric type
functions $_1 F_1 $, $_1 F_2 $ for bound states.

On the other hand, quasi-exactly solvable (QES) problems have
received great attention [10,11,12,13,14,15] in the last two
decades. The separation of variables for the Hamilton-Jacobi and
Schr\"{o}dinger equations in multiple orthogonal coordinate systems
is one of the chief problem in quantum mechanics, such as isotropic
harmonic oscillator in three dimensions [16,17,18].

In the past few years, quantum Hamilton-Jacobi (QHJ) formalism
[19,20,21,22] has been revisited. This formalism, formulated as a
theory related to the classical transformation theory \cite{20}, was
proposed by Leacock and Padgett in 1983 and they also formulated QHJ
method in one and three dimensions for a spherical potential which
is separable \cite{19}. The application of QHJ to a class of
solvable potentials has been re-explored in one dimensional space in
great detail by Bhalla, Kapoor and collaborators \cite{21}. It is
shown that QHJ formalism is a useful method to obtain eigenvalues
and eigenfunctions of the Hamiltonian system, for both ES and QES
problems [22]. Sextic oscillator and circular potential are
discussed by Geojo and collaborators \cite{22} as an application of
the QES models.

The motivation of this article is to introduce two dimensional QHJ
approach as an adaptation of Leacock's work \cite{19} and to solve
two dimensional singular oscillator which is introduced first time
by Calogero-Sutherland \cite{23}. In ref. [6], singular oscillator
in parabolic coordinates is solved within polynomial expansion and
solutions are obtained in a general form. In this study, as a first
application of QHJ approach to singular oscillator, solutions will
be obtained both in cartesian and parabolic coordinates.
Furthermore, we aim to generate new quasi exactly solvable
potentials by using Infeld-Hull factorization method \cite{24} that
has been developed recently by Dong \cite{25}.

Singular oscillator as the super-integrable potential is given as
\cite{6}
\begin{equation}
 V_{1}(x,y)=\frac{1}{2}
\omega^{2}(4x^{2}+y^{2})+k_{1}x+\frac{k^{2}_{2}-\frac{1}{4}}{2y^{2}}
\label{1}
\end{equation}
where $k_{2} > 0$. It is stated in [6] that for $k_2 > 1/2$, the
motion is restricted on one of the half planes $(-\infty < x <
\infty ,\;  y > 0)$ or $(-\infty < x < \infty, \; y < 0)$, whereas
for $0 < k_{2} < 1/2$ in whole plane $(x, y)$.

In the present work, in section II, two dimensional quantum
Hamilton-Jacobi formalism is introduced and exact solutions are
obtained for the singular oscillator in cartesian coordinates.
Section III involves QHJ formalism and QES solutions of singular
oscillator in parabolic coordinates. In section IV Infeld-Hull
factorization method has been used to generate new QES potentials.
Conclusions are given in the last section.

There are several formulations of quantum mechanics  such as
Heisenberg's matrix, Schr\"{o}dinger wave function, Feynman's path
integral, Wigner's Phase Space, density matrix, second quantization,
variational, de Broglie-Bohm pilot wave and Hamilton Jacobi
formulations. Quantum Hamilton-Jacobi formalism originally comes
from the classical Hamilton Jacobi theory which was first formulated
in the form of present work by Leacock and Padgett in 1983
\cite{19}. Then this formulation was investigated in a different
approach by Bhalla, Kapoor and their collaborators [21,22]. Let us
write

\begin{equation}
\overrightarrow p = \overrightarrow \nabla W \label{2}
\end{equation}
where $W$ is the quantum Hamilton characteristic function related to
the solution of the Schr\"{o}dinger equation

\begin{equation}
 - \frac{\hbar ^2}{2m}\nabla ^2\psi (x,y) + \left(V(x,y)-E \right)\,\psi (x,y)
 =0
 \label{3}
\end{equation}
by the definition $\psi (x,y) = \exp ({\frac{iW}{\hbar }})$. Using
this definition of wave function in Schr\"{o}dinger equation gives

\begin{equation}
( {\overrightarrow \nabla W})^2 - i\hbar \overrightarrow \nabla .(
{\overrightarrow \nabla W}) = 2m\; (E - V(x,y)). \label{4}
\end{equation}
If Eq.(\ref{2}) is used in Eq.(\ref{4}) the QHJ equation is obtained
as

\begin{equation}
{\overrightarrow p } ^2 - i \hbar \overrightarrow \nabla .
{\overrightarrow p } = 2m\; (E - V(x,y)), \label{5}
\end{equation}
in other words $\overrightarrow p $ is the logarithmic derivative of
$\psi (x,y)$ which can be written in the form of

\begin{eqnarray}
\overrightarrow p = - i\hbar \frac{\overrightarrow \nabla \psi
(x,y)}{\psi (x,y)} \; . \label{6}
\end{eqnarray}
Eq.(\ref{5}) can be separated by using quantum Hamilton
characteristic function $W(x,y) = W(x,\lambda _1 ) + W(y,\lambda _2
)$. Here $\lambda _1 $ and $\lambda _2 $ correspond to two cartesian
separation constants, $E=\lambda_{1}+\lambda_{2}$. The quantum
momentum functions (QMF) corresponding to the system are given as

\begin{equation}
p_x = \frac{\partial W_x }{\partial x}, \quad p_y = \frac{\partial
W_y }{\partial y} \label{7}
\end{equation}
from now onwards we set $\hbar = 2m = 1$. Using Eq.(\ref{7}) in
(\ref{5}), QHJ equation can be separated as follows

\begin{equation}
p_x ^2 - i\partial _x p_x - \lambda _1 + 2\omega ^2 x^2 + k_1 x = 0
\label{8}
\end{equation}

\begin{equation}
p_y^2 - i\partial _y p_y - \lambda _2 + \frac{\omega ^2}{2}y^2 +
\frac{k_2 ^2 - 1 / 4}{2y^2} = 0 \; . \label{9}
\end{equation}
To solve Eq.(\ref{8}) within the QHJ formalism, let us change the
variable as $z = x + \frac{k_1 }{4\omega ^2}$. Hence, Eq.(\ref{8})
turns into the form of
\begin{equation}
p_z ^2(z) - i\frac{\partial p_z }{\partial z} - (\varepsilon _1 -
2\omega ^2z^2) = 0 \label{10}
\end{equation}
where $\varepsilon _1 = \lambda _1 + \frac{k_1^{2}}{8\omega ^2} \;$.
In Eq.(\ref{10}), QMF $p_z (z,\varepsilon _1 )=p_z (z)$ has n poles
which are the zeros of the wave function. As it is seen from the
Eq.(\ref{10}), the residue at each of these poles is $ - i$. This
QMF has not any other poles except at infinity. In the limit of
large $z$, QMF approaches to $p_z (z,\varepsilon _1 ) \approx \pm
i\sqrt 2 \omega z$. Then we can write
\begin{equation}
p_z (z,\varepsilon _1 ) \approx \pm i\sqrt 2 \omega z + Q(z)
\label{11}
\end{equation}
where $Q(z)$ coming from the Liouville theorem is to be determined
later. The correct sign of $p_z (z,\varepsilon _1 )$ can be obtained
by using the condition of square integrability of the wave function.
The wave function is given as $\psi (z) = e{ }^{i\int {p_z
(z,\varepsilon _1 )dz} }$ for a bounded wave function at large $z$,
we choose the positive sign of QMF in Eq.(\ref{11}). Then one can
write the QMF function as
\begin{equation}
p_z (z,\varepsilon _1 ) = \sum\limits_{k = 1}^n { - \frac{i}{z - z_k
}} + i\sqrt 2 \omega z + \phi (z), \label{12}
\end{equation}
where $\phi (z)$ is analytic and bounded at infinity. From
Liouville's theorem [19,20,21,22], $\phi (z)$ has to be a constant
which is zero for $z\rightarrow \infty$. In Eq.(\ref{12})
$\sum\limits_{k = 1}^n {\frac{ 1}{z - z_k }} =
\frac{P^{\prime}(z)}{P(z)}$, where $P(z) = \prod\limits_{k = 1}^n
{(z - z_k )} $. Substitution of Eq.(\ref{12}) in Eq.(\ref{10})
yields
\begin{equation}
P^{\prime \prime}(z) - 2\sqrt 2 \omega zP^{\prime}(z) + (\varepsilon
_1 - \sqrt 2 \omega )P(z) = 0 \label{13}
\end{equation}
the solution of this equation gives the Hermite polynomials $P_n (z)
= H_n (\sqrt{\sqrt 2 \omega} z)$ and $\lambda _1 $ can be obtained
as $\lambda _1 = (2n+1)\sqrt{2}\omega - \frac{k_1^{2} }{8\omega
^2}$. The wave function has the following form
\begin{eqnarray}
 \psi (z) &=& \exp (i\int {p_z\;dz} ) \nonumber\\
 &=& B_n \;\exp [ - \frac{\sqrt{2}}{2}\omega z^{2}] \;H_n [\sqrt{\sqrt 2
\omega} z]. \nonumber
\end{eqnarray}
where $B_n$ is a normalization constant. Now let us consider the
solutions of Eq.(9); there is a singularity at $y=0$ in
Eq.(\ref{9}), then, one can expand
$p_{y}=\frac{b_{1}}{y}+a_{0}+a_{1}y+a_{2}y^{2}+...$ and use in
Eq.(\ref{9}), therefore following quantities can be found:
$a_{1}=\pm \frac{i \omega}{\sqrt{2}}$ , \,\ $a_{0}=0$ and
$b_{1}=-\frac{i}{2}\pm \frac{i
\sqrt{2}}{2}\sqrt{k_{2}^{2}+\frac{1}{4}}$.\,\ If we expand $p_{y}$
as
\begin{equation}
p_{y}=\frac{b_{1}}{y}+\frac{i\omega}{\sqrt{2}}y-i\frac{P^{'}}{P}+C
\label{14}
\end{equation}
where $P$ is an $n$th degree polynomial, prime denotes the
derivative in terms of $y$ and $C$ appears from Liouville theorem
[19,20,21,22] which equals to zero for large $y$. By using
Eq.(\ref{14}) in Eq.(\ref{9}), the following form can be obtained:
\begin{equation}
P_{n}^{''}(y)+\left(\frac{2ib_{1}}{y}-\omega\sqrt{2}y\right)P_{n}^{'}(y)+\left(\lambda_{2}-
\left(\frac{1}{2}+ib_{1}\right)\omega \sqrt{2}\right)P_{n}(y)=0.
\label{15}
\end{equation}
So we get the solutions of this equation as
\begin{equation}
\lambda_{2}=\omega \sqrt{2} \left[(2n+1) \mp
\frac{\sqrt{2}}{2}\sqrt{k^{2}_{2}+\frac{1}{4}} \right] \label{16}
\end{equation}
and
\begin{equation}
P_{n}(y)=\frac{n!}{(ib_{1}-\frac{1}{2})_{n}}
L^{(ib_{1}-\frac{1}{2})}_{n}\left(\frac{\omega}{\sqrt{2}}y^{2}\right)
\nonumber
\end{equation}
where $L^{a}_{n}(y)$ stands for the associate Laguerre functions and
$(a)_{n}:=a(a+1)...(a+n-1)$ for $n> 0,\\
(a)_{0}=1$ \cite{24}. Using $P_{n}(y)$, the wave function is
obtained as
\begin{equation}
\psi_{n}(y)=N \frac{n!}{(ib_{1}-\frac{1}{2})_{n}} y^{ib_{1}}
 \exp{(-\frac{\omega}{2\sqrt{2}}y^{2})}
 L^{(ib_{1}-\frac{1}{2})}_{n}\left(\frac{\omega}{\sqrt{2}}y^{2}\right) \label{17}
\end{equation}
here $N$ is a normalization constant. Using
\begin{equation}
 b_{1}=-\frac{i}{2}\pm \frac{i
\sqrt{2}}{2}\sqrt{\frac{1}{4}+k^{2}_{2}}  \label{18}
\end{equation}
in Eq.(\ref{17}), then wave function becomes
\begin{equation}
\psi_{n}(y)=N \frac{n!}{\left(\mp
\frac{\sqrt{2}}{2}\sqrt{k^{2}_{2}+\frac{1}{4}}\right)_{n}}y^{\frac{1}{2}(1\pm
\sqrt{2} \sqrt{k^{2}_{2}+ \frac{1}{4}} )}
 \exp{(-\frac{\omega}{2\sqrt{2}}y^{2})}
  L^{(\mp
\frac{\sqrt{2}}{2}\sqrt{k^{2}_{2}+
\frac{1}{4}})}_{n}\left(\frac{\omega}{\sqrt{2}}y^{2}\right).
\label{19}
\end{equation}
When $b_{1}$ has a positive sign, then $k_{2}< \frac{1}{2}$. As a
special case; for the value of $k_{2}=\frac{1}{2}$, positive sign of
$b_{1}$ should be taken and in this case Laguerre functions can be
written in terms of Hermite polynomials by using the relations
\cite{24}
\begin{equation}
   H_{2n}(y)= (-1)^{n}2^{2n}n! L^{-1/2}_{n}(y^{2}) \;, \quad \;
  H_{2n+1}(y)= (-1)^{n}2^{2n+1}n!y L^{-1/2}_{n}(y^{2}).
  \label{20}
\end{equation}
From the values of $\lambda_{1}$ and $\lambda_{2}$, the total energy
of the system can be written as
\begin{equation}
  E_{n}=\sqrt{2}\omega(2n+2)-\frac{k_{1}^{2}}{8\omega^{2}}\mp\frac{\sqrt{2}\omega}{2}
  \sqrt{k^{2}_{2}+\frac{1}{4}}
  \nonumber
\end{equation}
where $\lambda_{1}$ and $\lambda_{2}$ are labeled with $n_{1}$ and
$n_{2}$, $n=n_{1}+n_{2}$. The results agree with \cite{6}, the
difference of $\sqrt{2}$ arises from choosing the separation
constants.

Parabolic coordinates $\xi$ and $\eta$ are connected with the
cartesian $x$ and $y$ by  $x=\frac{1}{2}(\xi^{2}-\eta^{2})$ and
$y=\xi \eta$ respectively, for $\xi \in \textbf{R}$, \,\ $\eta < 0$.
Then, the potential relation in Eq.(\ref{1}) turns into the
following form
\begin{equation}
V(\xi,
\eta)=\frac{\omega^{2}}{2}\left(\xi^{4}+\eta^{4}-\xi^{2}\eta^{2}\right)+
\frac{1}{2}k_{1}\left(\xi^{2}-
\eta^{2}\right)+\left(\frac{k^{2}_{2}-\frac{1}{4}}{2}\right)\frac{1}{\xi^{2}\eta^{2}}
\label{21}
\end{equation}
Schr\"{o}dinger equation in parabolic coordinates is given as
\begin{equation}
 - \frac{\hbar ^2}{2m}\nabla ^2\psi (\xi,\eta) +
  \left(V(\xi,\eta)-E \right)\,\psi (\xi,\eta)=0
 \nonumber
\end{equation}
where $\nabla ^2=\frac{1}{\xi^{2}+
\eta^{2}}(\frac{\partial^{2}}{\partial
\xi^{2}}+\frac{\partial^{2}}{\partial \eta^{2}})$. By using the
definition $\psi (\xi,\eta) = \exp ({\frac{iW}{\hbar }})$ in
Schr\"{o}dinger equation gives
\begin{equation}
( {\overrightarrow \nabla W})^2 - i\hbar \overrightarrow \nabla .(
{\overrightarrow \nabla W}) = 2m\; (E - V(\xi,\eta)). \nonumber
\end{equation}
The logarithmic derivative of $\psi (\xi,\eta)$ gives the quantum
momentum function $\overrightarrow p = - i\hbar
\frac{\overrightarrow \nabla \psi (\xi,\eta)}{\psi (\xi,\eta)}$.
From now on we set $\hbar = 2m = 1$. Then, quantum characteristic
function $W(\xi,\eta)$ and quantum momentum functions are defined
below
\begin{eqnarray}
W(\xi,\eta) = W(\xi,E_1 ) + W(\eta,E_2 ), \,\ p_\xi = \frac{\partial
W_\xi }{\partial \xi}, \quad p_\eta = \frac{\partial W_\eta
}{\partial \eta} \nonumber
\end{eqnarray}
Hence, QHJ equation in parabolic coordinates can be separated as
\begin{equation}
p^{2}_{\xi}-ip^{'}_{\xi}-E_{1}
\xi^{2}+\frac{\omega^{2}}{2}\xi^{6}+\frac{1}{2}k_{1}\xi^{4}+\frac{A}{\xi^{2}}-T=0
\label{22}
\end{equation}
\begin{equation}
p^{2}_{\eta}-ip^{'}_{\eta}-E_{2}
\eta^{2}+\frac{\omega^{2}}{2}\eta^{6}-\frac{1}{2}k_{1}\eta^{4}+\frac{A}{\eta^{2}}+T=0
\label{23}
\end{equation}
where $A=\frac{k^{2}_{2}-1/4}{2}$ and $T$ is a separation constant.
Let us solve Eq.(\ref{22}) and firstly expand $p_{\xi}$ in series as
\begin{equation}
p(\xi,E_{1})=\frac{b_{1}}{\xi}+a_{0}+a_{1}\xi+a_{2}\xi^{2}+a_{3}\xi^{3}
\label{24}
\end{equation}
and substitute in Eq.(\ref{22}), we obtain $a_{3}=\pm i\sqrt{B}$,
\,\ $a_{2}=0$, \,\ $a_{1}=\pm \frac{ik_{1}}{4\sqrt{B}}$, \,\
$a_{0}=\pm \sqrt{T}$ and $b_{1}=\frac{i}{2}(-1+\varepsilon \mu)$.
Here, $B=\frac{\omega^{2}}{2}$, \,\ $\varepsilon=\pm 1$ and
$\mu=\sqrt{1+4A}$. Now we expand $p_{\xi}$ in Laurent series as
\begin{equation}
p(\xi, E_{1})=\frac{b_{1}}{\xi}\pm \sqrt{T}+\frac{ik_{1}}{4\sqrt{B}}
\xi+i\sqrt{B} \xi^{3}-i\frac{P_{n}^{'}}{P_{n}}+C_{1} \label{25}
\end{equation}
where $P_{n}(\xi)$ is an $n$th degree polynomial which is unknown
and $C_{1}$ is a constant due to Liouville theorem. If we use
Eq.(\ref{25}) in Eq.(\ref{22}), for large values of $\xi$,
$P=\xi^{n}$ and this leads to $C_{1} =0$, $T=0$. This case also
leads to
\begin{equation}
E_{1}\rightarrow \sqrt{2}\omega(n+2)\pm
\frac{1}{\sqrt{2}}\sqrt{\frac{1}{4}+k^{2}_{2}}-\frac{k_{1}^{2}}{8\omega^{2}}
\label{26}
\end{equation}
in $\xi\rightarrow \infty$ limit. The main equation is obtained by
using Eq.(\ref{25}) in Eq.(\ref{22}):
\begin{equation}
\left[\frac{\partial^{2}}{\partial
\xi^{2}}-\left(\frac{k_{1}\xi}{2\sqrt{B}}+2\sqrt{B}\xi^{3}-
\frac{2ib_{1}}{\xi}\right)\frac{\partial}{\partial\xi}-\left(\left(3\sqrt{B}-
\frac{k^{2}_{1}}{16B}
+2ib_{1}\sqrt{B}-E_{1} \right)\xi^{2}+\frac{k_{1}}{4\sqrt{B}}+
\frac{ib_{1}k_{1}}{2\sqrt{B}}\right)\right]P_{M}(\xi^{2})=0.
\label{27}
\end{equation}
where $P_{M}(\xi^{2})$ denotes the polynomials of degree $M$. With
this form, this equation can not be solved conventionally given in
[22]. The difference in our solution arises from the parameter
transformations. We use
\begin{equation}
E_{1}=3\sqrt{B}-\frac{k^{2}_{1}}{16B}+2ib_{1}\sqrt{B}+4\sqrt{B}M
 \label{28}
\end{equation}
in Eq.(\ref{27}) for notational reasons. Thus, this equation becomes
\begin{equation}
\left[-\partial_{\xi,\xi}+\left(\frac{k_{1}\xi}{2\sqrt{B}}+2\sqrt{B}\xi^{3}-
\frac{2ib_{1}}{\xi}\right)\partial_{\xi}-4\sqrt{B}M\xi^{2}-\epsilon+
\frac{k_{1}}{4\sqrt{B}}(1+3ib_{1})\right]P_{M}(\xi^{2})=0
 \label{29}
\end{equation}
where $\epsilon=\frac{ik_{1}b_{1}}{4\sqrt{B}}$. The simplest
solutions of Eq.(\ref{29}) will be introduced.
\\
Case 1: Now let us find one explicit solution which occurs when
$M=0$. In this case, the polynomial
\begin{equation}
P_{0}(\xi^{2})=1
 \label{30}
\end{equation}
is a constant. Then,
\begin{equation}
\epsilon^{+}= \frac{k_{1}}{2\omega}\left(\frac{5}{2\sqrt{2}}-
\frac{3}{4}\sqrt{1+4k^{2}_{2}}\right) \label{31}
\end{equation}
is obtained for the positive sign of the residue $b_{1}$. This also
leads to $k_{2}=\frac{\sqrt{7}}{2}$. Corresponding wave function is
given as
\begin{equation}
\psi^{+}=\xi^{\frac{1}{2}-\frac{1}{2\sqrt{2}}\sqrt{1+4k^{2}_{2}}}
e^{-\left(\frac{k_{1}}{4\sqrt{2} \omega}\xi^{2}+\frac{\omega
}{4\sqrt{2}}\xi^{4} \right)} \label{32}
\end{equation}
Here, the boundary conditions require
$\frac{1}{2\sqrt{2}}\sqrt{1+4k^{2}_{2}} < \frac{1}{2}$, but using
$k_{2}=\frac{\sqrt{7}}{2}$, it is seen that the solutions are not
physical for the wave function. And the solutions for the negative
sign of the residue $b_{1}$ are
\begin{equation}
\epsilon^{-}= \frac{k_{1}}{2\omega}\left(\frac{5}{2\sqrt{2}}+
\frac{3}{4}\sqrt{1+4k^{2}_{2}}\right) \label{33}
\end{equation}
that leads to $k_{2}=\frac{1}{2}$ and
\begin{equation}
\psi^{-}=\xi^{\frac{1}{2}+\frac{1}{2\sqrt{2}}\sqrt{1+4k^{2}_{2}}}
e^{-\left(\frac{k_{1}}{4\sqrt{2} \omega}\xi^{2}+\frac{\omega
}{4\sqrt{2}}\xi^{4} \right)}. \label{34}
\end{equation}
\\
Case 2: $M=1$ stands for the solutions in two dimensional space
which requires
\begin{equation}
P_{1}(\xi^{2})=\xi^{2}+s \label{35}
\end{equation}
where $s$ is a constant. Substituting this into Eq.(\ref{29}), one
obtains
\begin{equation}
\epsilon_{1,2}=\frac{3}{4\sqrt{B}}(1+ib_{1}k_{1}) \mp
\sqrt{8\sqrt{B}+16ib_{1}\sqrt{B}+\frac{k^{2}_{1}}{4B}} \label{36}
\end{equation}
Again we can consider the solutions according to signs of the
residue. For the positive sign of residue, solutions are obtained as
\begin{equation}
\epsilon^{+}=\frac{3}{2\sqrt{2}\omega}\left(\frac{3}{2}-
\frac{1}{2\sqrt{2}}\sqrt{1+4k^{2}_{2}}\right)\mp \gamma \label{37}
\end{equation}

\begin{equation}
\psi^{+}=\xi^{\frac{1}{2}-\frac{1}{2\sqrt{2}}\sqrt{1+4k^{2}_{2}}}
\left(\left(k_{1}\pm
\omega\sqrt{2}\gamma\right)\xi^{2}-2\sqrt{2}\omega
\left(2-\frac{1}{\sqrt{2}}\sqrt{1+4k^{2}_{2}}\right)\right)
e^{-\left(\frac{k_{1}}{4\sqrt{2} \omega}\xi^{2}+\frac{\omega
}{4\sqrt{2}}\xi^{4} \right)} \label{38}
\end{equation}
where $\gamma=\sqrt{8\sqrt{2}\omega-4\omega
\sqrt{1+4k^{2}_{2}}+\frac{k^{2}_{1}}{2\omega^{2}}}$ and
$\frac{1}{2\sqrt{2}}\sqrt{1+4k^{2}_{2}} < \frac{1}{2}$. Solutions
for the negative sign of the residue are introduced as
\begin{equation}
\epsilon^{-}=\frac{3}{2\sqrt{2}\omega}\left(\frac{3}{2}+
\frac{1}{2\sqrt{2}}\sqrt{1+4k^{2}_{2}}\right)\mp \gamma \label{39}
\end{equation}
\begin{equation}
\psi^{-}=\xi^{\frac{1}{2}+\frac{1}{2\sqrt{2}}\sqrt{1+4k^{2}_{2}}}
\left(\left(k_{1}\pm
\omega\sqrt{2}\gamma\right)\xi^{2}-2\sqrt{2}\omega
\left(2+\frac{1}{\sqrt{2}}\sqrt{1+4k^{2}_{2}}\right)\right)
e^{-\left(\frac{k_{1}}{4\sqrt{2} \omega}\xi^{2}+\frac{\omega
}{4\sqrt{2}}\xi^{4} \right)} \label{40}
\end{equation}

where $\gamma=\sqrt{8\sqrt{2}\omega+4\omega
\sqrt{1+4k^{2}_{2}}+\frac{k^{2}_{1}}{2\omega^{2}}}$ and
$\frac{1}{2}+\frac{1}{2\sqrt{2}}\sqrt{1+4k^{2}_{2}} > 0$.

One can transform Eq.(\ref{29}) into a form
\begin{equation}
f^{''}(\xi^{2})+\left(\left(3\sqrt{B}-\frac{k^{2}_{1}}{16B}-2ib_{1}\sqrt{B}+
4\sqrt{B}M\right)\xi^{2}-B\xi^{6}-\frac{k_{1}}{2}\xi^{4}+\epsilon^{'}\right)f(\xi^{2})=0
\label{41}
\end{equation}
by using
\begin{equation}
P(\xi^{2})=e^{(\frac{\sqrt{B}}{4}\xi^{4}+\frac{k_{1}}{8\sqrt{B}}\xi^{2})}f(\xi^{2})
\label{42}
\end{equation}
where $\epsilon^{'}=
-\epsilon+\frac{7ik_{1}b_{1}}{4\sqrt{B}}+\frac{k_{1}}{2\sqrt{B}}$.
One can see that Eq.(\ref{41}) is in the form of Schr\"{o}dinger
equation with a potential
\begin{equation}
V(\xi)=B\xi^{6}+\frac{k_{1}}{2}\xi^{4}-\left(3\sqrt{B}-\frac{k^{2}_{1}}{16B}-2ib_{1}\sqrt{B}+
4\sqrt{B}M\right)\xi^{2} \label{43}
\end{equation}
For large values of $M$, solutions of Eq.(\ref{41}) can be obtained
but an ansatz for the $f$ which is more compact can be written as
\cite{15}
\begin{equation}
f=\prod^{M}_{i=1} (\frac{\xi^{2}}{2}-\eta_{i}) \xi^{\varsigma}
e^{-(\frac{\sqrt{B}}{4}\xi^{4}+\frac{k_{1}}{8\sqrt{B}}\xi^{2})}
\label{44}
\end{equation}
where $\sqrt{\eta_{i}}$ are the zeros of the $f(\xi^{2})$ and
$\varsigma=\frac{k^{2}_{1}}{32B^{3/2}}-ib_{1}+\frac{3}{2}$. Then,
$\epsilon^{'}$ is obtained by using Eq.(\ref{44}) in Eq.(\ref{41})
as
\begin{equation}
\epsilon^{'}=(2M+1)\sqrt{2}\omega +\frac{k^{2}_{1}}{8\sqrt{2}\omega}
\pm \frac{\omega}{2}\sqrt{1+4k^{2}_{2}}- 4\sqrt{2}\omega
\sum^{M}_{i=1}\eta_{i}.  \label{45}
\end{equation}
Following the same procedure, the solutions of Eq.(\ref{23}) can be
obtained as in the previous way.

Now we introduce Infeld-Hull factorization method firstly [25,26].
For a given QES $V(\xi)$ potential, Schr\"{o}dinger equation is [15]
\begin{equation}
\psi^{''}_{j}(\xi)=(V(\xi)-E_{j})\psi_{j}.  \label{46}
\end{equation}
Using following logarithmic derivative in Schr\"{o}dinger equation
\begin{equation}
y_{j}(\xi)=\frac{\psi^{'}_{j}(\xi)}{\psi_{j}(\xi)}  \label{47}
\end{equation}
gives Riccati type equation
\begin{equation}
V(\xi)-E_{j}=y_{j}^{'}(\xi)+y^{2}_{j}(\xi).  \label{48}
\end{equation}
For $M+1$ solutions Schr\"{o}dinger equation is
\begin{equation}
\psi^{''}_{2k+j}(\xi)=(V(\xi)-E_{2k+j})\psi_{2k+j}  \label{49}
\end{equation}
$k=0,1,...M$ [15] and from Eq.(\ref{46}), Eq.(\ref{47}) and
Eq.(\ref{48}) we get
\begin{equation}
-\frac{\partial^{2}}{\partial \xi^{2}}+y^{'}_{j}+y^{2}_{j}=
(y_{j}+\frac{\partial}{\partial \xi})(y_{j}-\frac{\partial}{\partial
\xi}). \label{50}
\end{equation}
If the operator $y_{j}(\xi)-\frac{\partial}{\partial \xi}$ acts on
\begin{equation}
(y_{j}+\frac{\partial}{\partial \xi})(y_{j}-\frac{\partial}{\partial
\xi})\psi_{2k+j}=(E_{2k+j}-E_{j})\psi_{2k+j} \label{51}
\end{equation}
then, the Schr\"{o}dinger equation with a new potential and wave
function is obtained [15]:
\begin{equation}
(-\frac{\partial^{2}}{\partial
\xi^{2}}+{\tilde{V}(\xi)})\tilde{\psi}_{2k+j}(\xi)=E_{2k+j}\tilde{\psi}_{2k+j}(\xi)
\label{52}
\end{equation}
where
\begin{equation}
\tilde{V}(\xi)=V(\xi)-2\frac{\partial^{2}}{\partial\xi^{2}}ln\psi_{j}(\xi)
\label{53}
\end{equation}

\begin{equation}
\tilde{\psi}=(E_{2k+j}-E_{j})
\frac{\int^{\xi}_{-\infty}\psi_{2k+j}\psi_{j}d\xi}{\psi_{j}}.
\label{54}
\end{equation}
From Eq.(53) it seems that new potential is generated by Darboux
type transformation from $V(\xi)$ [27,28,29]. In the language of
supersymmetric quantum mechanics $V(\xi)$ and $\tilde{V}(\xi)$ are
known as supersymmetric partner potentials. Now new potentials can
be obtained for $M=0$ and $M=1$ cases using Eq.(\ref{53}). For the
case of $M=0$, new potential is
\begin{equation}
\tilde{V}=\frac{1}{2}k_{1}\xi^{4}+
\left(-E+3\sqrt{2}\omega+\frac{k^{2}_{1}}{4\sqrt{2}\omega}\right)\xi^{2}+
\frac{\left(\frac{k^{2}_{2}+\frac{7}{4}}{2}\pm
\frac{1}{\sqrt{2}}\sqrt{1+4k^{2}_{2}}\right)}{\xi^{2}}+
\frac{k_{1}\left(\frac{3}{2}\pm
\frac{1}{\sqrt{2}}\sqrt{1+4k^{2}_{2}}\right)}{\sqrt{2}\omega}
 \label{55}
\end{equation}
where $\pm$ arises from positive and negative signs of the residue
values. For $M=1$, the potential has a more complex form which is
\begin{equation}
\tilde{V}=\frac{2a^{2}\omega^{2}}{(a\xi^{2}+b)^{2}}\xi^{10}+
\frac{2ab(1+\omega^{2})+\frac{3k_{1}a^{2}}{2}}{(a\xi^{2}+b)^{2}}\xi^{8}+
\eta_{6}(\xi)\xi^{6}+\eta_{4}(\xi)\xi^{4}+\eta_{2}(\xi)\xi^{2}+
\frac{\eta_{-2}}{\xi^{2}}-\eta_{01}(\xi)+\eta_{02}(\xi)
 \label{56}
\end{equation}
where
\begin{equation}
\eta_{6}(\xi)=\frac{\omega^{2}}{2}+
\frac{2(b\sqrt{B}+\frac{k_{1}a}{4\sqrt{B}})^{2}}{(a\xi^{2}+b)^{2}}-
\frac{4a^{2}(2+p)\sqrt{B}}{(a\xi^{2}+b)^{2}},
 \label{57}
\end{equation}

\begin{equation}
\eta_{4}(\xi)=\frac{k_{1}}{2}+\frac{2a\sqrt{B}(2p+7)-
\frac{k^{2}_{1}}{8B}-\frac{k_{1}b}{2}}{a\xi^{2}+b}-
\frac{4abp+4(\frac{k_{1}a}{4\sqrt{B}}+b\sqrt{B})(a(2+p)-
\frac{k_{1}b}{4\sqrt{B}})}{(a\xi^{2}+b)^{2}},
 \label{58}
\end{equation}

\begin{equation}
\eta_{2}(\xi)=-E-\frac{(-\frac{k_{1}a(2p+5)}{2\sqrt{B}}+
\frac{k^{2}_{1}}{8B}-2\sqrt{B}bp(2+\frac{3}{p}))}{a\xi^{2}+b}+
\frac{2a^{2}(2+p-\frac{k_{1}b}{4\sqrt{B}})^{2}-2(b^{2}p\sqrt{B}+
\frac{k_{1}abp}{4\sqrt{B}}+b^{2}p\sqrt{B})}{(a\xi^{2}+b)^{2}},
 \label{59}
\end{equation}

\begin{equation}
\eta_{-2}(\xi)=\frac{k^{2}_{2}-\frac{1}{4}}{2}-\frac{2bp(p-1)}{a\xi^{2}+b}+
\frac{2b^{2}p^{2}}{(a\xi^{2}+b)^{2}},
 \label{60}
\end{equation}

\begin{equation}
\eta_{01}(\xi)=\frac{2a(2+p)(1+p)-\frac{k_{1}b(1+p)}{2\sqrt{B}}}{a\xi^{2}+b},
 \label{61}
\end{equation}

\begin{equation}
\eta_{02}(\xi)=\frac{4(abp(2+p)-\frac{k_{1}b^{2}p}{4\sqrt{B}})}{(a\xi^{2}+b)^{2}},
 \label{62}
\end{equation}

where $a=k_{1}\pm \omega\sqrt{2}$, $b=-2\sqrt{2}\omega(2\pm
\frac{1}{\sqrt{2}}\sqrt{1+4k^{2}_{2}})$, $p=\frac{1}{2}\pm
\frac{1}{2\sqrt{2}}$.

As a first application of two dimensional QHJ approach, we
investigate the solutions of $2D$ singular oscillator potential
which admits separation of variables in two different coordinate
systems. After separation of variable, this potential can be handled
as two different one-dimensional systems. Then, by following one
dimensional QHJ approach of Bhalla and collaborators work, we get
the exact solutions in terms of Hermite and Laguerre functions for
cartesian coordinates which are consistent with the solutions in
\cite{30,6}. In section III, it is shown that the present method
\cite{22} is problematic for obtaining polynomial solutions in
parabolic coordinates. Hence, we use transformations of parameters
in order to obtain solutions for $M=0,1$ cases, thus only the
limited number of solutions are obtained in parabolic coordinates.
The transformation of Eq.(\ref{29}) leads to a Hamiltonian with a
new potential and solutions of this form are obtained in more
compact form. Despite the more general results obtained in Ref.[6]
by using polynomial solutions, it is stated that a limited number of
wave function solutions can be obtained correctly by using QHJ
approach for singular oscillator in parabolic coordinates. Thereof,
it can be an open problem to investigate a generalization of the
approach for the QES problems.

Finally, new quasi-exactly solvable potentials  are constructed for
$M=0$ and $M=1$ cases by using Infeld-Hull factorization method. By
means of this method the solutions of more complicated potential can
be found just by using the solution of singular oscillator potential
from Eq.(54), so there is no need to solve the corresponding
Schr\"{o}dinger or quantum Hamilton-Jacobi equation.

This work was supported in part by the Scientific and Technical
Research Council of Turkey (T\"{U}B\.{I}TAK).



\begin{references}

\bibitem{1} Bender C M , Wang Q 2001 J. Phys. A: Math. Gen. 34 9835.

\bibitem{2}  Bagchi B,  Gorain P S, Quesne C 2006 Mod. Phys. Lett. A 21
2703; Dong S H, Sun G H, Cassou M L 2005 Phys. Lett. A  340 94.

\bibitem{3}  Znojil M, J. Phys. A: Math. Gen. 2006 39 4047 ; Mustafa O,
 Mazharimousavi S H 2006 Phys.Lett. A 357  295.

\bibitem{4}   Cannata F,  Ioffe M V,  Nishnianidze D N 2006 Theor.Math.Phys. 148
 960.

\bibitem{5} Cang Z M, Bang W Z 2005 Chin. Phys. Lett.
22 2994.

\bibitem{6}  Kalnins E G,  Miller W Jr.,  Pogosyan G S 2006
J. Math. Phys. 47 033502.

\bibitem{7} Flugge 1971 Problems in Quantum Mechanics  (New York: Springer -
Verlag).


\bibitem{8}  de Lange O L and Raab R E 1991 Operator Methods in Quantum Mechanics
( Oxford: Claredon Press)


\bibitem{9} Natanzon G A 1979  Theor.Math.Phys. 38
146.

\bibitem{10} Turbiner A V 1988 Comm. Math. Phys. 118  467.

\bibitem{11} Razavy M 1980 Am. J. Phys. 48  285.

\bibitem{12} Gangopadhyaya A, Khare A, Sukhatme U P 1995 Phys. Lett. A 208
261.

\bibitem{13} Tkachuk V M  1998 Phys. Lett. A 245  177.

\bibitem{14}  Kuliy T V,   Tkachuk V M 1999 J. Phys. A: Math. Gen. 32 2157.

\bibitem{15} Ushveridze A G  1994 Quasi-Exactly Solvable Models in Quantum
Mechanics  (Taylor and Francis).

\bibitem{16}Eisenhart L P 1948 Phys.Rev. 74  87.

\bibitem{17}  Miller, W Jr. 1977 Symmetry and Separation of Variables (Rhode Island
Addison-Wesley Pub. C.)


\bibitem{18}  Kalnins E G 1986 Separation of Variables for Riemannian Spaces of Constant
 Curvature   Pitman ( Essex: Longman)

\bibitem{19} Leacock R A,  Padgett M J 1983 Phys. Rev. Lett  50  3; 1983 Phys.
Rev. D 28 2491.

\bibitem{20} Jordan P  1926 Z. Phys. 38  513;  Schwinger J  1958 Quantum Electrodynamics
(NY: Dover Publications).

\bibitem{21} Bhalla R S,  Kapoor A K, and Panigrahi P K 1997 Am. J. Phys. 65 1187
; Bhalla R S, Kapoor A K and Panigrahi P K 1997 Mod. Phys. Lett. A
12  295.

\bibitem{22}  Ranjani S S,  Kapoor A K and Panigrahi P K 2004 Mod. Phys. Lett. A 19
 2047; Ranjani S S,  Kapoor A K and Panigrahi P K 2005 Ann.
Phys.  320  164; Ranjani S S,  Kapoor A K  and Panigrahi P K 2005
Int. J. Theor. Phys. 44 1167;  Geojo K G, Ranjani S S,  Kapoor A K
2003 J. Phys A: Math. Gen. 36  4591;  Ranjani S S, Geojo K G, Kapoor
A K, Panigrahi P K 2004 Mod. Phys. Lett. A. Vol 19, No. 19 1457;
Ranjani S S,  Kapoor  A. K.; Panigrahi  P. K. 2005 Int. J. of Mod.
Phys. A 20  4067.

\bibitem{23}  Calogero F 1971  J. Math. Phys.  12  419;
 Sutherland B 1972 Phys. Rev. A  4  2019; 1969 J. Math. Phys. 12
2191;  Sutherland B 1970 J. Math. Phys  12  246.

\bibitem{24}  Arfken G B 1995 Mathematical Methods for Physicists (London:Academic Press)

\bibitem{25}  Infeld L,  Hull T E 1951 Rev. Mod. Phys.  23  21.

\bibitem{26}  Dong S H 2007 Factorization Method in Quantum Mechanics
(Springer, Kluwer Ac. Pub.)

\bibitem{27}  Junker G  1996 Supersymmetric Methods in Quantum and Statistical
Physics (Berlin: Springer).


\bibitem{28}  Cooper F,  Khare A and  Sukhatme U 1995 Phys. Rep. 251  267.

\bibitem{29}  Matveev V B,  Salle M A  1991 Darboux Transformations and
Solitons (Berlin: Springer).

\bibitem{30}  Perelomov A 1986 Generalized Coherent States and Their Applications
(Berlin: Springer)


\end{references}
\end{document}